\documentclass{emulateapj}

\def\sco{Sco~X-1}  

\begin{document} 
 
\title{A Hard X-ray View on Scorpius X-1 with INTEGRAL: non-Thermal
Emission ?} 
 
\author{T. Di Salvo\altaffilmark{1}, P. Goldoni\altaffilmark{2,3},
L. Stella\altaffilmark{4}, M. van der Klis\altaffilmark{5}, 
A. Bazzano\altaffilmark{6}, L. Burderi\altaffilmark{7}, 
R. Farinelli\altaffilmark{8}, F. Frontera\altaffilmark{8}, 
G.L. Israel\altaffilmark{4}, M\'endez\altaffilmark{9}, 
F. Mirabel\altaffilmark{10}, N.R. Robba\altaffilmark{1}, P. Sizun\altaffilmark{3}, 
P. Ubertini\altaffilmark{6}, W. H. G. Lewin\altaffilmark{11}
}
\altaffiltext{1}{Dipartimento di Scienze Fisiche ed Astronomiche, 
Universit\`a di Palermo, via Archirafi 36 - 90123 Palermo, Italy;
email:disalvo@fisica.unipa.it} 
\altaffiltext{2}{APC, Laboratoire Astroparticule et Cosmologie, UMR 7164, 
11 Place Marcelin Berthelot, 75231 Paris Cedex 05, France.}
\altaffiltext{3}{CEA Saclay, DSM/DAPNIA/Service d'Astrophysique, F91191, 
Gif-sur-Yvette France.}
\altaffiltext{4}{Osservatorio Astronomico di Roma, via Frascati 33, 
00040 Monteporzio Catone (Roma), Italy.} 
\altaffiltext{5}{Astronomical Institute "Anton Pannekoek," University of
Amsterdam and Center for High-Energy Astrophysics, Kruislaan 403, 
NL 1098 SJ Amsterdam, the Netherlands.}
\altaffiltext{6}{Istituto di Astrofisica Spaziale e Fisica Cosmica, Sezione di 
Roma, INAF, via Fosso del Cavaliere 100, I-00133 Rome, Italy.}
\altaffiltext{7}{Universit\`a degli Studi di Cagliari, Dipartimento
di Fisica, SP Monserrato-Sestu, KM 0.7, 09042 Monserrato, Italy.}
\altaffiltext{8}{Dipartimento di Fisica, Universit\'a di Ferrara, 
Via Paradiso 12, 44100 Ferrara, Italy.}
\altaffiltext{9}{SRON Netherlands Institute for Space Research, Sorbonnelaan 2, 
3584 CA Utrecht, Netherlands.}
\altaffiltext{10}{European Southern Observatory - Vitacura, Casilla 19001, Santiago 19
Chile.}
\altaffiltext{11}{Center for Space Research, Massachusetts Institute of Technology, 
77 Massachusetts Avenue, Cambridge, MA 02139-4307, USA.}
 
\begin{abstract} 
We present here simultaneous INTEGRAL/RXTE observations of Sco X-1, and 
in particular a study of the hard X-ray emission of the source and its
correlation with the position in the Z-track of the X-ray color-color
diagram. We find that the hard X-ray (above about 30 keV) emission of 
Sco X-1 is dominated by a power-law component with a photon index of
$\sim 3$. The flux in the power-law component slightly decreases
when the source moves in the color-color diagram in the sense of 
increasing inferred mass accretion rate from the horizontal branch to
the normal branch/flaring branch vertex. It becomes not significantly
detectable in the flaring branch, where its flux has decreased 
by about an order of magnitude. These results present close analogies
to the behavior of GX 17+2, one of so-called Sco-like Z sources.
Finally, the hard power law in the spectrum of Sco X-1 does not show
any evidence of a high energy cutoff up to $100 - 200$ keV, strongly 
suggesting a non-thermal origin of this component.
\end{abstract} 

\keywords{accretion discs -- stars: individual: \sco\ --- stars: neutron
stars --- X-ray: stars --- X-ray: general --- X-ray: binaries} 

 
\section{Introduction} 

Hard X-ray emission in the brightest low-mass X-ray binaries (hereafter LMXBs), 
the so-called Z-sources, was occasionally detected in the past
(see e.g.\  Peterson \& Jacobson 1966).
These results received relatively little 
attention, mostly because the lack of a broad-band spectral coverage
did not permit to establish whether an extra component was indeed required 
to fit the hard spectrum of these sources.
Renewed interest in the hard X-ray emission properties of bright LMXBs
was motivated by recent broad-band studies mainly performed with 
RXTE (2 -- 200 keV) and BeppoSAX (0.1 -- 200 keV). These have shown 
that most Z-sources display variable, hard power-law shaped components, 
dominating their spectra above $\sim 30$~keV (see Di Salvo \& Stella 2000
for a review).

The hard component detected in bright (otherwise soft) LMXBs can be fitted
by a power law, with photon index in the range 1.9--3.3, contributing from
1\% to 10\% of the observed ($0.1-200$ keV) source luminosity.
The presence of these components in Z sources seems sometimes to be related 
to the source state or its position in the X-ray color-color diagram 
(hereafter CD). The clearest example to date is in the BeppoSAX observation 
of GX~17+2, where the hard component (a power-law with photon index of 
$\sim 2.7$) showed the strongest intensity in the horizontal branch (HB) of 
its CD (Di Salvo et al.\ 2000). A factor of 20 decrease was observed when 
the source moved from the HB to the normal branch (NB), 
i.e.\ from low to high (inferred) mass accretion rate. 
A hard tail was also detected in almost all the currently known Z sources
(e.g. Di Salvo et al.\ 2001; Iaria et al.\ 2001; Di Salvo et al. 2002; 
Asai et al.\ 1994).
The fact that a
similar hard component has been observed in several Z sources indicates
that this is probably a common feature of these sources.
However, the origin of this hard component is still poorly understood.
While in most cases the hard component becomes weaker at higher
accretion rates, HEXTE observations of Sco X--1 showed a hard power-law 
tail in 5 out of 16 observations, without any clear correlation with the
position in the CD (D'Amico et al.\ 2001). The thermal vs.\ non-thermal 
nature of this component remains to be addressed, yielding important
information on the production mechanism.

Sco X--1, the brightest persistent X-ray source in the sky, is also
the brightest radio source among neutron star LMXBs, with a mean
radio flux about 10 times higher than that of the other Z sources (e.g.\ 
Fender \& Hendry 2000). 
A hard X-ray power-law component has been
observed in RXTE/HEXTE (20--200 keV) data of this source
(D'Amico et al.\ 2001). As already mentioned, contrary to the case of 
GX~17+2, in Sco X--1 the flux of this component was observed to vary without any 
clear correlation with the position in the CD. 
Interestingly, Strickman \& Barret (2000) report that the hard X-ray emission 
present in OSSE data of Sco X--1 may be correlated with periods of radio flaring.

To study the hard X-ray emission in \sco, the brightest of these sources, 
as well as its correlation with other source properties (such as radio 
emission and fast timing variability), we have performed a campaign of
observations of Sco X--1 with INTEGRAL and RXTE. Part of these observations
were also done simultaneously with radio VLBI observations (which will
be discussed elsewhere). 
The INTEGRAL spectrum of \sco\ shows with high statistical
significance the presence of a hard (power-law) component, without any
clear exponential cutoff up to $\sim 100-200$ keV. We also find clear
evidence that the intensity of this component is correlated with the 
position of the source in the X-ray CD.

\section{Observations and analysis} 

Sco X-1 was observed during two complete INTEGRAL revolutions on 2003 July 30
-- August 1 and 2003 August 11 -- 13.
The INTEGRAL payload consists of two main $\gamma$-ray instruments, a spectrometer, 
SPI (Vedrenne et al.\ 2003) and an imager, IBIS, and of two monitor instruments,
the X-ray monitor JEM-X (3 -- 35 keV, Lund et al.\ 2003) and the Optical Monitoring 
Camera (V band, 500-600 nm, Mas-Hesse et al.\ 2003). 
The IBIS instrument (Ubertini et al.\ 2003) covers the energy range
between 20 keV and 8 MeV with two detectors, ISGRI (Lebrun et al.\ 2003)
and PICsIT (Di Cocco et al.\ 2003) and has a field of view of $29^\circ
\times 29^\circ$ at half sensitivity ($9^\circ \times 9^\circ$ fully
coded) with a point spread function of $12'$ FWHM. The SPI also 
covers the energy range between 20 keV and 8 MeV with a FOV of $31^\circ$ 
diameter ($16^\circ$ fully coded) and an angular resolution of $2.5^\circ$.

Observations were performed in the usual INTEGRAL format, i.e. the
observation was split in separate exposures (``science windows'', or 
scws), each lasting $\sim 3600$ sec, followed by a 5 minutes slew. 
The exposures were arranged following a $5 \times 5$ dither pattern.
The IBIS and SPI instruments were operated in standard mode during the
whole observation, while JEM-X was in a non-standard mode (SPEC) which,
unfortunately, is not yet calibrated. We therefore discarded the JEM-X data
from our analysis. The effective exposure time was 372 ks for IBIS and
358 ks for SPI\footnote{Due to the much lower statistics of the SPI data, 
we will concentrate here on the analysis of ISGRI data of \sco.}. 
The IBIS data were analyzed using the standard analysis procedures of OSA version
5 and the latest response  matrices (October 2004) rebinned to 26 channels
between 15 and 800 keV and the latest spectral extraction routines
(Goldwurm et al.\ 2003). 
A systematic error of $1\%$ was applied to all the INTEGRAL/ISGRI spectra.


During all the observations, Sco X-1 was the only source detected
in the wide FOV of the instruments. The source coordinates as
derived from the ISGRI mosaic image in the 20 -- 35 keV energy band
are: RA = 16 19 54.9, DEC = --15 38 34.4 (uncertainty $\pm 10\arcsec$, 
1 sigma confidence level) at about $10\arcsec$ from
the SIMBAD position (McNamara et al. 2003). 

To study the source spectral state in the standard X-ray band,
we also analyzed data from the Proportional Counter
Array (PCA; Zhang et al.\ 1993) on board RXTE, which consists of five
co-aligned Proportional Counter Units (PCUs), with a total collecting area of
6250 cm$^2$ and a field of view, limited by collimators, of $1^\circ$ FWHM,
sensitive in the energy range $2-60$ keV.  The RXTE observation, performed
simultaneously with the INTEGRAL observation, is divided into two parts of 
175 ks each, approximately two weeks apart from each other.
We selected intervals for which the elevation angle of the source above 
the Earth limb was greater than 10 degrees. On a few occasions, some 
of the five PCUs were off; we therefore used only data from PCUs 2 and 3
which where on for most of the observation.
The X-ray CD of Sco X-1 during the INTEGRAL observations, 
obtained from the PCA data, is shown in Figure~1. During each
of the two observations the source described a fairly complete Z-track 
in the CD.

In order to check for the presence of the hard tail in our data, we first analysed 
the INTEGRAL (IBIS/ISGRI, energy band 20 keV -- 200 keV) spectrum 
integrated over the whole observation (see Figure~2, left panel).
Similar to what has been done for other LMXBs of the Z-class, we fitted
the INTEGRAL spectrum with the Comptonization model {\tt compTT} 
(whose description is given in Titarchuk 1994).  {\tt Comptt} is an 
analytical model describing the thermal Comptonization of soft photons
(for which a Wien spectrum is assumed) inverse-Compton scattered in a hot
electron cloud with optical depth $\tau$ and whose temperature can range
from a few keV to 500 keV. This model includes relativistic effects and 
works for optically thick and optically thin regimes. The geometry of the
Comptonizing cloud can be either spherical or disk-like. In this paper we
assume a spherical geometry (this assumption only affects the value of the
optical depth derived from the fit).

This model gives a good fit of the soft part of the \sco\ spectrum up to
$\sim 40-50$ keV. Above this energy, a hard excess is clearly visible in
the residuals, independently of the particular Comptonization model used to fit 
the soft part of the spectrum. The fit is significantly improved by adding
to the {\tt compTT} model a power-law with photon index $\sim 3.1$ 
(this gives a reduction of the $\chi^2$/dof from 892/13 to 14.4/11).
We tested the presence of a thermal cutoff in the hard power-law;
substituting the power law with a cutoff power-law (that is a power law
multiplied by an exponential cutoff) does not improve the 
fit significantly (the latter model gives a $\chi^2$/dof = 15.6/10), and
the temperature of the exponential cutoff is $k T > 200$ keV ($90\%$
confidence level).
The best-fit parameters for the ISGRI (20 -- 200 keV) 
spectrum are reported in Table~1; data and residuals with respect
to the best fit model are shown in Figure~2 (left panel).

To look for variability in the hard component with the spectral state
of the source, as measured by its position in the X-ray CD, we divided 
the Z-track in the CD of \sco\ into four parts corresponding to
the HB/upper-NB, the NB, the flaring branch (FB), and the NB/FB vertex, 
respectively. We therefore extracted
INTEGRAL/ISGRI spectra for each of the time intervals mentioned above, 
resulting in four CD resolved spectra. 
Unfortunately there was no superposition between RXTE and INTEGRAL data 
in the FB during the first part of the observation. 
The INTEGRAL/ISGRI exposure times for the four intervals were 52.4 ks, 
45.5 ks, 60.6 ks and 6 ks. 
We fitted each of this spectra with {\tt comptt} and a cutoff power law.
This model gave a good fit of the first three spectra, with little variability
in the spectral parameters (see Table~1).
For three of these spectra, the hard power law component was required in 
order to fit the data.
Remarkably the FB spectrum did not require a power-law
component and could be fitted with a simple Comptonization model
(see Fig.~2, right panel).
Including the power law in the spectral fit of the FB 
spectrum, with the photon index fixed at 3.1, a good fit requires a 
decrease of the normalization of the power law component by a factor
at least 5 with respect to the average spectrum. In Table~1 we 
show the total X-ray flux of \sco\ calculated in the $20-40$ keV
and in the $40-200$ keV energy range, Flux (20--40) and 
Flux (40--200), respectively. While the Flux (20--40) decreases
from the HB to the NB/FB vertex 
and then increases when the source goes to the FB, 
the Flux (40--200) always decreases, varying by about one order of 
magnitude when the source moves from the HB to the FB.

Although from Table~1 there seems to be a clear trend of the 
$40-200$ keV flux to decrease from the HB to the 
NB and FB vertex, the uncertainties on the 
hard X-ray flux are still quite large to draw a firm conclusion.
In fact, in order to calculate the flux in a given energy range
and the associated uncertainty, XSPEC needs to use the total 
best-fit model, which includes both the {\tt comptt} and the 
cutoff power-law component; each of these components has several
parameters and the uncertainties on all the parameters are taken into 
account in the calculation of the source flux and its uncertainty.
If we are interested in knowing the flux of the hard power-law 
component alone in a given energy range (to see how this component 
evolves when the source moves in the X-ray CD), it is more convenient 
to use the model named {\tt pegpwrlw} in XSPEC. This model allows to 
directly calculate the power-law flux in a given energy range 
($20-200$ keV is our choice) and the associated uncertainty.
We therefore substituted the cutoff power-law model with the 
{\tt pegpwrlw} model to fit the INTEGRAL spectra of \sco.
The results of the fits with the {\tt comptt} plus {\tt pegpwrlw} model 
are also reported in Table~1 (since the parameters of the 
{\tt comptt} did not change significantly these are shown only once
in the table). The decrease of the power-law $20-200$ keV flux 
along the CD is evident. For clarity, we have also
plotted these power-law fluxes in Figure~1 (inset).

\section{Discussion} 

We report on a spectral analysis of a simultaneous 
$\sim 300$~ks-long observation of \sco\ with INTEGRAL and RXTE.
We show that the addition of a hard power-law component dominating
the X-ray spectrum above $\sim 30$ keV proves necessary for a good
fit of the INTEGRAL ($20-200$ keV) spectrum. 
The power law is quite steep, with a photon index of about 3.1,
contributing up to 12\% of the observed $20 - 200$ keV luminosity, and
does not show any evidence of a high energy cutoff up to $100-200$ keV.
Similarly to what was observed in the BeppoSAX observation of the
Z source GX~17+2, the presence of the hard component in \sco\ seems
to be related to the position of the source in the Z track; in fact 
we observe a clear trend of the $20-200$ keV power-law flux to decrease 
when the inferred mass accretion rate increases (i.e.\ from the HB
to the NB/FB vertex). At the highest 
inferred mass accretion rate (i.e.\ in the FB of the X-ray
CD), the hard X-ray emission seems to disappear completely (the hard
X-ray flux decreases by about one order of magnitude with respect to the
flux measured in the HB/upper-NB). Note, however, that 
the behavior of the hard component in the FB may be variable 
(c.f.\ D'Amico et. al.\ 2001).

One of the most interesting results of this analysis is that the 
hard power law detected in \sco\ does not show any evidence of a 
high energy cutoff, with lower limits on the electron temperature 
that are in most cases above 100 keV (the lower limit on a thermal
cutoff in the \sco\ averaged INTEGRAL spectrum, where the statistics 
is the highest, is about 200 keV). Such high temperatures are
not expected in a system like \sco, because of the strong Compton 
cooling due to the primary soft spectrum, where most of the energy
is emitted. These results therefore indicate that, in analogy with
the steep hard power-law emission detected in some soft states 
(e.g.\ intermediate and very high) of systems containing a black hole 
candidates, which do not show any energy cutoff up to a few hundreds
keV, the hard power law observed in Z sources may be of non-thermal origin.

The most probable origin of these components are therefore Comptonization in 
a hybrid thermal/non-thermal corona (where the Maxwellian velocity distribution 
of the electrons have a non-thermal high-velocity tail, e.g.\ Poutanen \&
Coppi 1998; see Farinelli et al.\ 2005 and D'A\'\i\ et al.\ 2006 for a 
successful application of this model to the case of GX~17+2 and \sco, 
respectively). For instance, it is possible that a hard population of 
electrons is accelerated in internal shocks at the base of a jet. Otherwise 
the hard power law may originate in a mildly relativistic ($v/c \ga 0.5$) 
bulk motion of matter close to the compact object. 
Therefore non-thermal Comptonization in outflows (or jets) may be the origin 
of these components, with flatter power laws corresponding to higher optical 
depths of the scattering medium and/or higher bulk electron velocities 
(e.g.\ Psaltis 2001). 

It has been proposed that non-thermal, high energy 
electrons, responsible for the hard tails observed in Z sources, might be
accelerated in a jet (Di Salvo et al.\ 2000). This is also confirmed by
the observed correlation between the strength of the hard X-ray emission and
the position of the source in the X-ray CD. In fact all Z sources are detected
as variable radio sources with the highest radio flux associated with
the HB  and the lowest with the FB (Migliari \&
Fender 2006, and references therein). In other words, both the radio emission 
from these objects, which is probably due to jets (Fender \& Hendry 2000), and
the hard X-ray emission are anticorrelated with the inferred mass accretion rate. 
In this respect, it will be very important to study the direct correlation
between the presence and strength of the hard X-ray emission and the radio
activity of the source. The radio (VLA) observations of \sco, simultaneous to 
part of the INTEGRAL observations, is under investigation and will be 
discussed elsewhere.

\acknowledgements 
This work was partially supported by 
the Ministero della Istruzione, della Universit\`a e della Ricerca (MIUR),
and by the Agenzia Spaziale Italiana (ASI). 
 


\begin{table}
\footnotesize
\caption{Results of the fitting of the \sco\ INTEGRAL/ISGRI (20--200 keV) 
spectra. }
\label{table:1}
\newcommand{\m}{\hphantom{$-$}}
\newcommand{\cc}[1]{\multicolumn{1}{c}{#1}}
\renewcommand{\tabcolsep}{0.6pc} 
\renewcommand{\arraystretch}{1.2} 
\begin{center}
\begin{tabular}{@{}llllll}
\hline
Parameter  &     Averaged      & HB/UNB & NB &	NB/FB & FB \\
\hline
Comptt + Cutoff Power-law Model          & 	&	&	&	 \\
\hline
$k T_0$ (keV) & $1.6$ (frozen) & $1.6$ (frozen) & $1.6$ (frozen) &
				$1.6$ (frozen) & $1.6$ (frozen) \\
$k T_{\rm e}$ (keV) & $3.31^{+0.02}_{-0.04}$ & $2.95^{+0.11}_{-0.02}$ & $3.43 \pm 0.02$ &
				$3.42^{+0.03}_{-0.27}$ & $3.14 \pm 0.20$ \\
$\tau$        & $5.70^{+0.10}_{-0.05}$ & $8.5^{+1.0}_{-0.9}$ & $5.35 \pm 0.11$ &
				$5.21 \pm 0.12$ & $6.5^{+1.7}_{-1.1}$ \\
PhoIndex      & $3.12 \pm 0.07$ & $3.11^{+0.17}_{-0.50}$ & $2.70^{+0.43}_{-0.93}$ &
				$3.20^{+0.21}_{-0.66}$ & $3.1$ (frozen) \\
$kT$ (keV)    & $> 220$         & $> 70$ & $> 160$ & $> 118$ & -- \\
Flux (20--40) & $6.1 \pm 0.2$ & $7.12 \pm 0.41$  & $5.85 \pm 0.37$    
		    & $4.43 \pm 0.50$  & $7.31 \pm 0.45$ \\
Flux (40--200) & $2.6 \pm 1.8$ & $3.7 \pm 1.6$ & $2.5 \pm 1.8$    
		    & $1.85 \pm 0.89$ & $0.36 \pm 0.36$ \\
$\Delta \chi^2$     & $878$            	    & $464$	& 	$164$
					    & $102$	&	$ - $ \\
$\chi^2 / d.o.f.$   & $15.6 / 10$      	    & $7.0 / 10$ & $10.8 / 10$ 
					    & $12.9 / 10$ & $7.5 / 12$ \\
\hline
Comptt + Pegpwrlw  Model             	& 	&	&	&	 \\
\hline
PhoIndex		& $3.31^{+0.08}_{-0.17}$	    & $3.29^{+0.33}_{-0.22}$ &	
	$3.04^{+0.32}_{-0.62}$   &  $3.59^{+0.65}_{-0.42}$  &  $3.1$ (frozen) \\
Flux (20--200)  	& $6.27 \pm 0.35$		    & $9.4^{+1.4}_{-0.8}$    &  
	$5.4^{+1.5}_{-0.4}$   &  $5.10^{+0.85}_{-0.63}$ & $< 0.93$ \\
$\chi^2(d.o.f.)$              & $14.4 / 11$ 		    & $6.8 / 11$             &  
		$11.3 / 11$   &  $13.0 / 11$ &  $7.5 / 12$  \\
\hline
\end{tabular}\\[2pt]
\end{center}
\footnotesize
The model consists of a Comptonized spectrum modeled by {\tt comptt}, and
a cutoff power law or the {\tt Pegpwrlw} model to fit the hard component. 
$k T_0$ is the temperature of the seed photon (Wien) spectrum, $k T_e$ the
electron temperature and $\tau$ the optical depth in a spherical geometry.
Fluxes (20--40) and (40--200) are the total flux from the source in units of 
$10^{-9}$ erg cm$^{-2}$ s$^{-1}$ in the 20 -- 40 keV energy range and in units 
of $10^{-10}$ erg cm$^{-2}$ s$^{-1}$ in the 40 -- 200 keV energy range, respectively. 
On the other hand, Flux (20--200) is the power-law flux in the 
20--200 keV range in units of $10^{-10}$ erg cm$^{-2}$ s$^{-1}$. 
$\Delta \chi^2$ is the variation of $\chi^2$ for the addition
of a power law to the model. Note that the addition of a power law does not
change the $\chi^2$ for the FB spectrum. $\chi^2 / d.o.f.$ is the 
final reduced $\chi^2$ for the best-fit model.
All the uncertainties are calculated at 90\% confidence level, upper limits at
95\% confidence level.
\end{table}


\begin{figure}
\includegraphics[width=7cm,height=8cm]{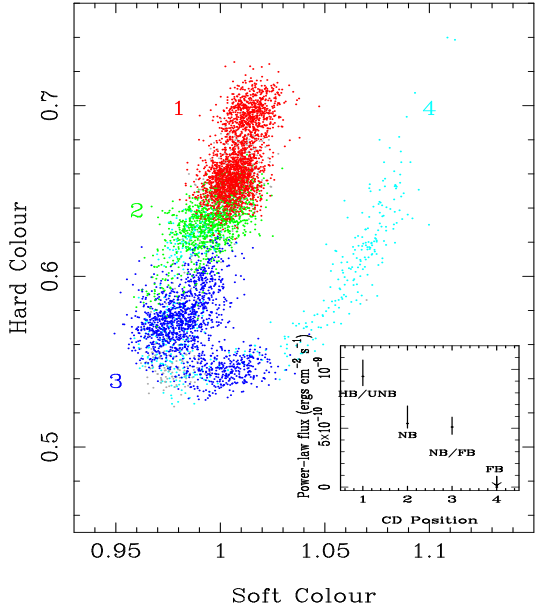}
\caption{	
Color-color diagram of Sco X--1 from PCA data during the
simultaneous INTEGRAL/RXTE observation. Here Soft Color is the ratio
of the count rate in the energy bands [$3.5-6$ keV]/[$2-3.5$ keV] and
Hard Color is the ratio [$9.7-16$ keV]/[$6-9.7$ keV], respectively.
To take into account possible small gain changes during the RXTE observations,
the Hard and Soft colors of \sco\ are normalized with respect to
the colors (calculated in the same energy range) obtained from RXTE observations 
of Crab acquired close to the dates of our observations.
Only data form PCUs 2 and 3, which were on for most of the observation, 
have been used. The four colors indicate the four regions (also indicated 
with numbers from 1 to 4) in which the CD has been divided, 
from which INTEGRAL spectra were extracted. 
{\bf Inset:} The measured power-law flux in the $20-200$~keV range plotted
for each of the four CD-resolved spectra.  \label{fig1}}
\end{figure}


\begin{figure}
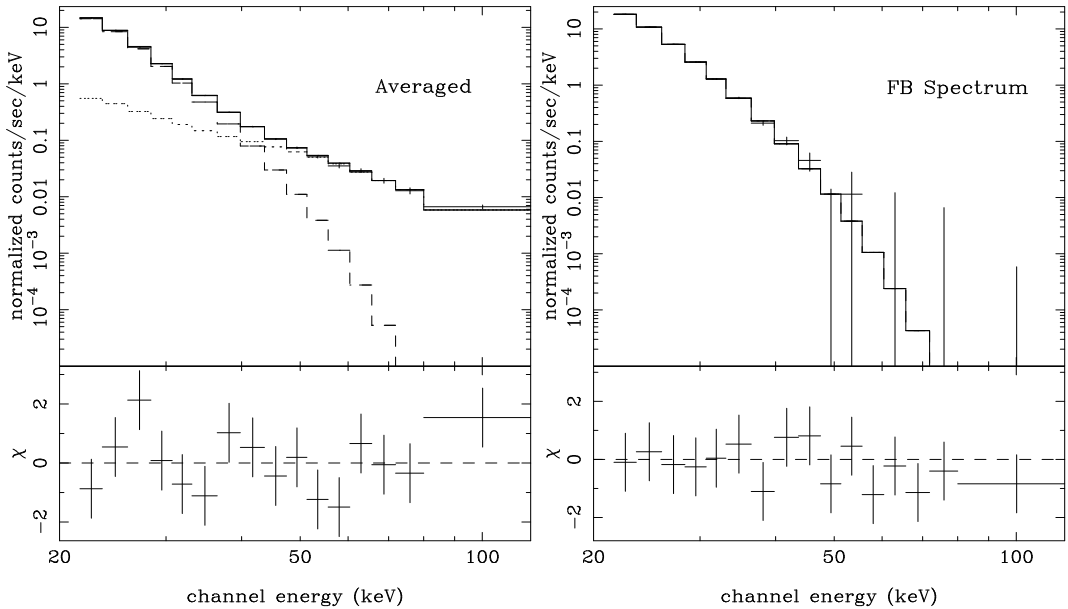

\includegraphics[width=7cm,height=8cm]{f2a.eps}
\includegraphics[width=7cm,height=8cm]{f2b.eps}
\caption{{\bf Left:} ISGRI (20 -- 200 keV) 
averaged spectrum (left) and FB spectrum (right) of \sco\ 
together with the best-fit model (solid line on top of the data) composed 
of the Comptonization model ({\tt comptt}, dashed line) and a cutoff power 
law (dotted line). {\bf Right:} residuals in units of $\sigma$ with respect to 
the best fit models shown in Table~1.  
\label{fig2}}
\end{figure}


\begin{thebibliography}{} 
\bibitem[]{575} Asai, K., Dotani, T., Mitsuda, K., et al.\ 1994, PASJ, 46, 479
\bibitem[]{576} D'A\'\i, A., et al.\ 2006, ApJ, submitted
\bibitem[]{577} D'Amico, F., Heindl, W.A., Rothschild, R.E., Gruber, D.E. 2001, 
ApJ 547, L147 
\bibitem[]{579} Di Cocco, G., Caroli, E., Celesti, E., et al. 2003, A\&A, 411, L189
\bibitem[]{580} Di Salvo, T., Stella, L., Robba, N.R., et al. 2000, ApJ 544, L119
\bibitem[]{581} Di Salvo, T., \& Stella, L. 2000, Proceedings of the XXXVIIth 
Rencontres de Moriond, Eds. A. Goldwurm, D. Neumann, J. Tran Thanh Van, p. 67
\bibitem[]{583} Di Salvo, T., Robba, N.R., Iaria, R., et al. 2001, ApJ, 554, 49
\bibitem[]{584} Di Salvo, T., Farinelli, R., Burderi, L., et al. 2002, A\&A, 386, 535
\bibitem[]{585} Farinelli, R., et al.\ 2005, A\&A, 434, 25
\bibitem[]{586} Fender, R. P., \& Hendry M. A. 2000, MNRAS, 317, 1
\bibitem[]{587} Fomalont, E.~B., Geldzahler, B.~J., \& Bradshaw, C.~F. 2001a ApJ,
553, L27  
\bibitem[]{589} Fomalont, E.~B., Geldzahler, B.~J., \& Bradshaw, C.~F. 2001b ApJ,  
558, 283  
\bibitem[]{593} Goldwurm, A., David, P., Foschini, L., et al. 2003, A\&A, 411, L223
\bibitem[]{594} Iaria, R., Burderi, L., Di Salvo, T., et al. 2001, ApJ, 547, 412
\bibitem[]{597} Lebrun, F., Leray, J. P., Lavocat, P., et al. 2003, A\&A, 411, L141
\bibitem[]{598} Lund, N., Budtz-Joergensen, C., Westergaard, N. L., et al. 2003, 
A\&A, 411, L231
\bibitem[]{600} McNamara, B. J., et al. 2003, AJ, 125, 1437
\bibitem[]{601} Mas-Hesse, J. M., Gimenez, A., Culhane, J. L., et al. 2003, A\&A, 
411, L261
\bibitem[]{603} Migliari, S., \& Fender, R. P., 2006, MNRAS, in press
\bibitem[]{604} Peterson, L. E., \& Jacobson, A. S., 1966, ApJ, 145, 962 
\bibitem[]{606} Poutanen, J., \& Coppi, P. S., 1998, Phys. Scripta, T77, 57
\bibitem[]{607} Psaltis, D., 2001, ApJ, 555, 786
\bibitem[]{608} Strickman, M., \& Barret, D. 2000, Proc. of the fifth Compton 
Symposium, Eds. M. L. McConnell and J.M. Ryan, AIP Conference Proceedings, 
Vol. 510, p. 222
\bibitem[]{611} Titarchuk, L. 1994, ApJ, 434, 570
\bibitem[]{612} Ubertini, P., Lebrun, F., Di Cocco, G., et al. 2003, A\&A, 411, L131
\bibitem[]{613} Vedrenne, G., Roques, J.-P., Sch{\"o}nfelder, V., et al. 2003, 
A\&A, 411, L63
\bibitem[]{617} Zhang, W., Giles, A.B., Jahoda, K., Soong, Y., Swank, J.H., Morgan, 
E.H. 1993, Proc. SPIE 2006, 324
\end{thebibliography}
\end{document}